\documentclass[11pt]{article}
\usepackage[T1]{fontenc}
\usepackage{scr/aas_macros}
\usepackage{scr/mathStyle}
\usepackage[us,12hr]{datetime}
\pdfoutput=1
\usepackage{fancyhdr}
\usepackage{amsmath,amssymb,amsthm,latexsym,bbm,calc,wasysym,mathtools,empheq, yfonts,upgreek}
\usepackage[english]{babel}
\usepackage[usenames,dvipsnames]{xcolor}
\usepackage[english]{babel}
\usepackage{graphicx}
\usepackage[framemethod=tikz]{mdframed}
\usepackage[english]{babel}
\usepackage[utf8]{inputenc}
\usepackage{graphicx}
\usepackage{xcolor}
\usepackage{mathrsfs}
\usepackage{amsfonts}
\newcommand{\comment}[1]{}
\renewcommand{\baselinestretch}{1.11}
\renewcommand\footnoterule{{\color{pc1}\rule {8.15cm}{0.4pt}}}

\definecolor{rindou1}{rgb}{0.4431,0.2862,0.7960}
\definecolor{rindou2}{rgb}{0.0078,0.1215,0.4392}
\definecolor{lapis}{rgb}{0.0.0470,0.2941,0.5568}
\definecolor{mn}{rgb}{0.15, 0.35, 0.95}

\begin{document}

\definecolor{mycolor}{rgb}{0.122, 0.435, 0.698}
\definecolor{pc1}{rgb}{0.69, 0.25, 0.21}
\renewcommand{\baselinestretch}{1.08}
\renewcommand\footnoterule{{\color{pc1}\hrule height 0.7pt}}
\definecolor{green2}{cmyk}{0, 1, 0.5, 0}
\definecolor{lightgreen}{cmyk}{0.2, 0, 0.2, 0.2}
\definecolor{lightgray}{cmyk}{0.1,0.2,0,0.1}
\definecolor{lightgray2}{cmyk}{0.4,0.4,0,0.8}
\definecolor{black}{cmyk}{1.0,1.0,1.0,1.0}
\definecolor{brown}{rgb}{0.79, 0.16, 0.16}
\definecolor{cerulean}{rgb}{0.0, 0.48, 0.65}
\definecolor{blue2}{rgb}{0.0, 0.2, 0.4}
\definecolor{royalazure}{rgb}{0.4, 0.2, 0.92}
\definecolor{darkcerulean}{rgb}{0.03, 0.27, 0.89}
\definecolor{magenta1}{rgb}{0.89, 0.08, 0.48}
\definecolor{mc}{rgb}{0.122, 0.435, 0.898}
\definecolor{cb}{rgb}{0.29, 0.59, 0.82}
\definecolor{mn}{rgb}{0.05, 0.25, 0.65}

\hypersetup{%
	colorlinks,
	citecolor=magenta,
	filecolor=BrickRed,
	linkcolor=rindou2,
	urlcolor=mycolor,
	linktocpage=true
}

\title{\LARGE{\textsc{Tensor renormalization
of three-dimensional Potts model}}} 
\author{Raghav G. Jha}
\affiliation{Perimeter Institute for Theoretical Physics, 
Waterloo, Ontario N2L 2Y5, Canada}
\emailAdd{raghav.govind.jha@gmail.com}
\abstract{\\~\\ {\textsc{Abstract:} 
We study the $q$-state Potts models on a cubic lattice in the thermodynamic limit 
using tensor renormalization group transformations with the triad approximation. 
By computing the thermodynamic potentials, we locate
the first-order phase transitions for $10 < q  \le 20$ which has not been 
explored using any method. We also examine the efficiency of the triad approximation method 
in obtaining the fixed-point tensor and comment on how this can be improved. 
}}
\maketitle
\textbf{\label{sec:0}\section{Introduction}} 
The motivation of this work is to examine some recent developments in the tensor renormalization group method referred to as the `triad' method. We find that this method is reasonably precise in computing the potentials and critical temperatures for a range of models, including the $q$-state Potts model in three dimensions. However, this method most likely misses the correct fixed-point tensor. In order to reach this conclusion, we use an observable defined in Ref.~\cite{Gu_2009} and show that this method cannot accurately compute this 
either for the Ising model or for the $q$-state Potts model in three dimensions when we go sufficiently close to criticality for some fixed and finite bond dimension. Therefore, while this is a promising method to study statistical systems in higher dimensions, admittedly, much work remains to be done to generate a more precise scheme. 

The framework of renormalization group (or rather `semi-group') is probably the most important advance
in the last fifty years and has taught us the correct way to think about quantum field theories. One of the most exciting areas of research in the past fifteen years is the development that we can use tensors to effectively carry out the renormalization group (RG) procedure through some well-defined transformations numerically. In his famous paper on the solution of Kondo problem, Wilson emphasized and clearly pointed out that this feature (which was called the `fourth aspect' in his paper) 
of studying renormalization group theory on digital computers was the most exciting. 
In the last five decades with improved computing resources and the possibility of quantum computers in the future, this is even more exciting than Wilson had envisaged. This approach has taught us valuable 
things about entanglement, holography, phase transitions in statistical models while much more still remains
to be understood.  The numerical method introduced by Wilson for the solution of the impurity problem 
was not applicable to a wider class of one-dimensional models. The numerical RG program was stagnant until 1992, when White showed that by using the density matrices and making reasonable assumptions, 
it was possible to study the one-dimensional systems much more efficiently than any previous methods. 
The next development in the real-space renormalization group came when it was proposed that one can use multi-index objects called tensors to carry out RG transformations for gapped systems. The idea of tensor renormalization group (TRG) started with the seminal work of Levin and Nave in 2007 \cite{Levin:2006jai} and was later improved by a new method called `second renormalization group' (SRG) introduced in Ref.~\cite{Xie:2009zzd} which took the effect of the tensor environment into account to carry out a well-defined renormalization group flow.  However, both TRG and SRG were shown to be inefficient at the critical point. This gave rise to the idea of another well-known algorithm known as tensor network renormalization (TNR) \cite{Evenbly2015} which works better than both TRG and SRG for gapless systems and the convergence in the bond dimension is exponential rather than polynomial. Though all these methods work reasonably well for two-dimensional classical models, the extension to three or more dimensions is not straightforward. In this pursuit, a new approach called higher-order TRG (HOTRG) was introduced in Ref.~\cite{Xie2012}. However, the memory cost of this method scale rather poorly with dimensions as $D^{4d-2}$ and becomes unmanageable (unless one has access to supercomputers with such memory) quickly. To deal with this problem, a few years ago, a new approximate method called `triad TRG' (TTRG) was introduced. This method is basically the application of the `divide and conquer' rule applied to HOTRG. The triad RG significantly cuts down the memory requirements and computation time since a higher rank tensor is never explicitly made and offers a fairly acceptable level of accuracy by running on small machines especially in three and four dimensions. However, this method is certainly 
not as robust as HOTRG and the real effectiveness of this method and comparison to HOTRG in higher
dimensions remains a topic of discussion. For a recent review on tensor networks, we
refer the interested reader to Ref.~\cite{Meurice:2020pxc} and to Ref.~\cite{Efrati2014} 
for an extended review of real-space renormalization in statistical mechanics
over the past five decades. In this paper, we will focus on the triad method applied to the 
three-dimensional spin system especially the $q$-state Potts model and see how far we can go. This is a continuation of a long program of research where tensor networks have been applied extensively  
to spin systems and gauge theories in two and three dimensions~\cite{Liu:2013nsa, Wang2014, Bazavov:2019qih, Jha:2020oik, Bloch:2021mjw, Hostetler:2021uml}. Most of the tensor network studies of three-dimensional spin models are restricted to Ising model, however,  there was a study of $q=3$ Potts model in Ref.~\cite{Wang2014} and very recently we studied the $O(2)$ model using tensor methods in Ref.~\cite{Bloch:2021mjw}. 

The model we study in this paper was first defined\footnote{As noted in Ref.~\cite{Baxter1985}, Potts actually defined two separate models. The first was the $\mathbb{Z}_{N}$ model and assumed that at each site of a lattice there was a two-dimensional vector that could point in one of the $N$ equally spaced directions where the vectors interact with an energy proportional to their dot product. The second was the model which was defined on any graph and with edges joining pair of sites.} by Renfrey Potts in 1951 and describes a generalization of the Ising model. This is now known as 
$q$-state Potts model based on the local Hilbert space dimension of the individual spin. 
In fact, there was a previous study of models of this type already around 1943 
when the $q=4$ model was considered 
by Askhin-Teller (a model now known by their name). 
For a detailed review of Potts model, its history, and various exact 
solutions in different dimensions, we refer the interested reader to Ref.~\cite{Wu1982}. 
Though the model seems relatively straightforward, for the three-dimensional classical $q$-state Potts model, we do not even have an understanding of how the critical temperatures depend on $q$. Note that this is at least well-known on square lattice even though full solution does not exist for free energy for generic temperature with $ q \ge 3$ \cite{Baxter1985}. In two dimensions, 
by mapping to Ice-type models, this was computed to be: $ T_{c} = 1/\ln(1 + \sqrt{q})$. Though this reduces to 
Onsager result for $q=2$ (up to a multiplicative factor) and has been checked by Baxter and Hintermann for $ q \ge 4$, a rigorous proof still lacks for $q=3$, see Ref.~\cite{Wu1982} for details. 
In three dimensions, the determination of the critical temperature has to be done through numerical methods and even though a large volume of work has been done and numerical results are available up to $q = 10$, there is no work beyond this with any method! This is surprising for a model which was introduced more than 70 years ago. In this work, we aim to fill this gap by computing critical temperatures on cubic lattice for $10 < q \le 20$ using tensor network methods with the triad method developed in Ref.~\cite{Kadoh:2019kqk}.  

Another interesting study in the past for these models is the determination 
of the critical $q$ where the order of the phase transition changes and is referred to as $q_{c}$. 
This was determined to be $q_{c} = 4$ \cite{Baxter1985}, and $q_{c} = 2.35(5)$ \cite{Hartmann2005}
for two and three-dimensional models respectively. For $q \le q_{c}$, we have continuous (second-order) phase transition
which is the case for the Ising model in both two and three dimensions while for $q > q_{c}$ it is first-order. 
However, it is known that even for $q=3$ in three dimensions, it is still only weakly first-order. Our results confirm this behavior based on the observation that we cannot fit to the first-order ansatz well. The strength of the first-order transitions, measured by their latent heat, increases with $q$ and our result corroborates this behavior.

\begin{figure}
	\centering 
	\includegraphics[width=0.7\textwidth]{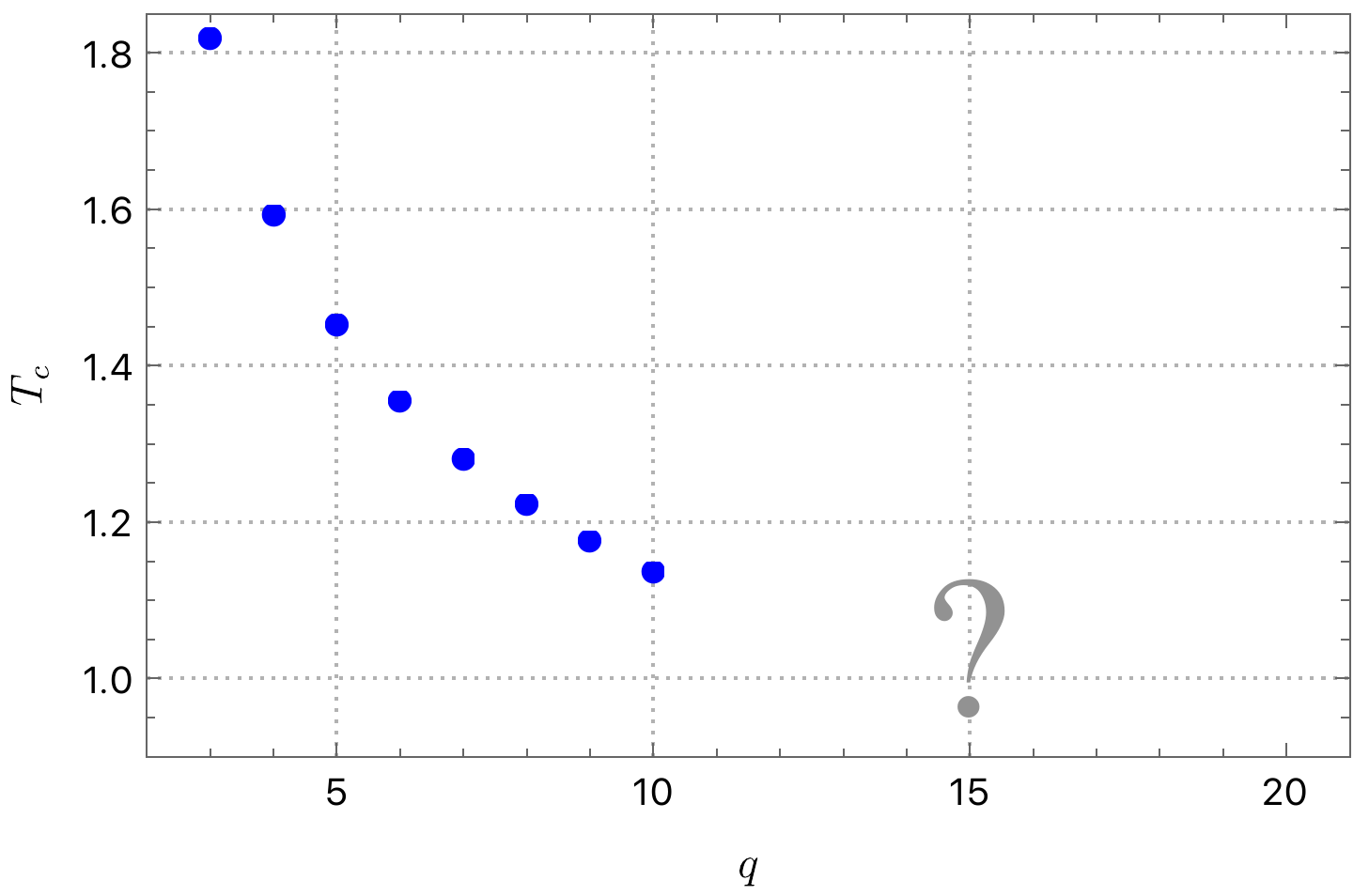}
	\caption{\label{fig:1}The critical temperature for $q$-state Potts model in three dimensions calculated using Monte Carlo methods. However, even after seventy years since this model was introduced, nothing is known for $q > 10$. The main purpose of this paper is to fill this gap. The results we obtain are given in Table~\ref{tab:dataMC}. It remains an open problem to find an expression for $T_{c}(q)$ for this model.}
\end{figure}
The outline of the paper is as follows. In Sec.~\ref{sec:1}, we introduce the model and write down the tensor
formulation in terms of triads. Then in Sec.~\ref{sec:2}, we start by checking some old results for $q=3$ obtained using HOTRG about eight years ago and for $q=10$ obtained using Monte Carlo methods. 
Then we present new results for $10 < q \le 20$. This is summarized in Fig.~\ref{fig:1}
We end the paper with a brief summary and future directions by mentioning
how the triad approximation still needs to be improved to accurately compute 
critical exponents in Sec.~\ref{sec:3}. It is clear from our work that though it fairly accurately computes 
potentials and can locate the critical temperatures, it does not capture the correct fixed-point tensor
behaviour around the critical points. 

\vspace{10mm} 
\section{\label{sec:1}Formulation and Observables}
\subsection{\label{subsec:1.1}$q$-state Potts model}

We consider the $q$-state Potts model with 
the Hamiltonian given by:
\begin{equation}
	\mathcal{H} = -J \sum_{\langle ij \rangle} \delta(\sigma_i,\sigma_{j}), 
\end{equation}
where $J$ is the coupling which we set equal to one, $\sigma_{i} = 1, \cdots q$ and  
$\langle ij \rangle$ are the nearest neighbours on a cubic lattice. The partition function is given by:
\begin{equation}
	Z = \sum_{\{\sigma_{i}\}} \prod_{\langle ij \rangle}  \exp\Big[\beta \delta(\sigma_i,\sigma_{j})  \Big],
\end{equation}
where $\beta$ is the inverse temperature. The tensor network approach is the representation of the 
path integral or partition function of a classical statistical system by a trace over 
some network. This can be schematically written as:
\begin{equation}
	Z = \int \mathcal{D}\phi~e^{-S} =  \sum_{\{\phi_{i}\}} \prod T_{\phi_i \phi_j \phi_k \phi_l \phi_m \phi_n}. 
\end{equation}
We can express this partition function as a tensor trace of network as:
\begin{equation}
	\label{eq:TP} 
	Z = \rm{tTr} \Big(\bigotimes T_{ijklmn} \Big). 
\end{equation}
It is well-known that the exact evaluation of (\ref{eq:TP}) is not possible and hence the 
goal is to approximate it to the best of our resources. The choice of different $T_{ijklmn}$ corresponds to 
different Lagrangian/partition function. This tensor description also gives rise to a renormalization group flow 
of tensor such that if we do it fairly accurately we obtain the correct fixed-point. However, complete removal of the short-distance correlations is difficult to achieve in practice and has so far not been worked out accurately in greater than two dimensions. 

\subsection{\label{subsec:1.2}Tensor description of the model}

We noted in the preceding subsection that the method of tensor networks approximates the partition function of a system as a tensor trace of some complicated network. In order to obtain this, we start by expressing the initial rank-six tensor in terms of triads and coarse-grain the system several times to reach the thermodynamic limit. There are several ways of constructing the initial tensor for this model \cite{Zhao_2010,genzor2020tensor} and they are all equivalent. We use the simplest one which 
\begin{figure} 
	\includegraphics[width=0.92\textwidth]{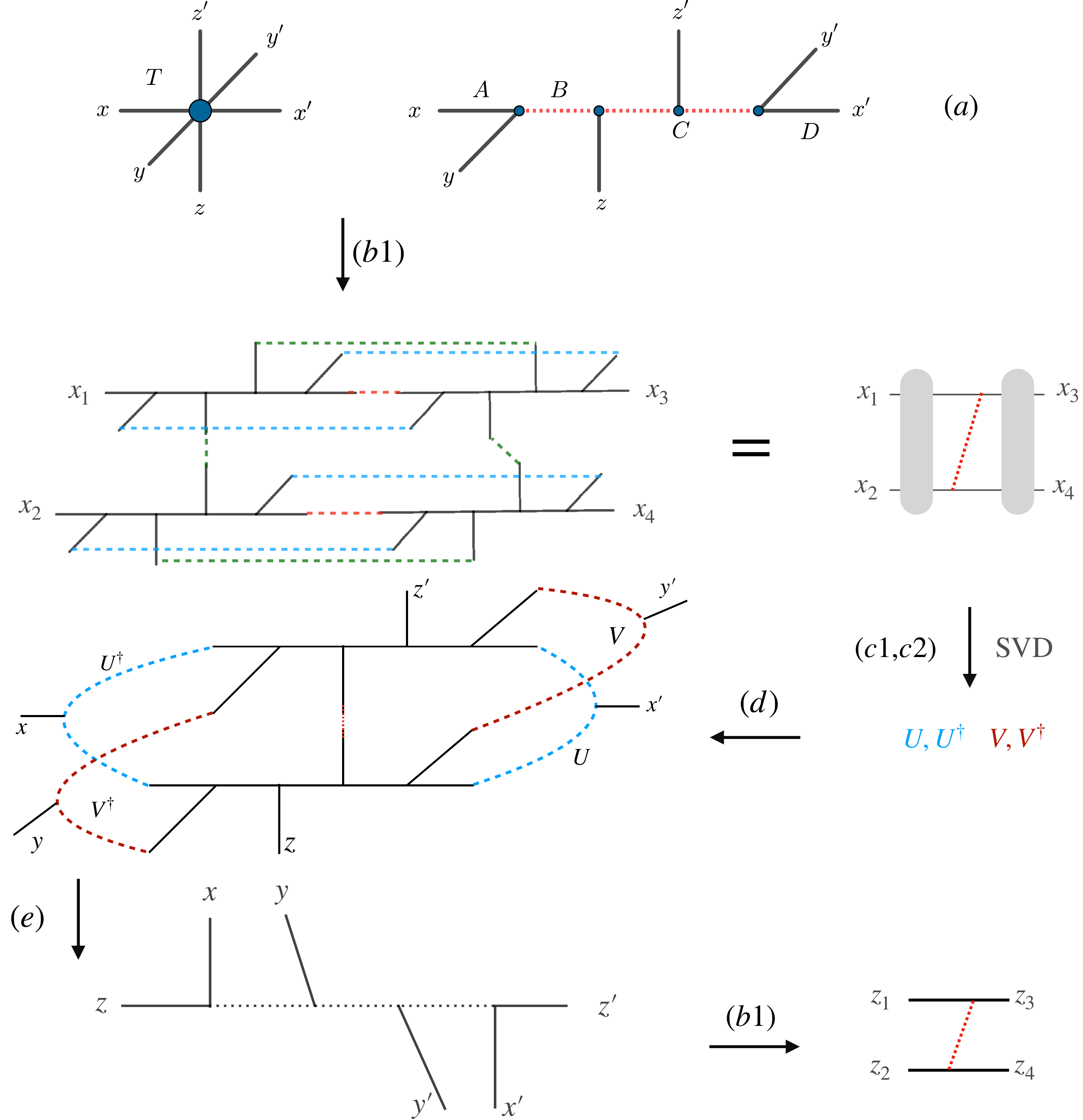}
	\caption{\label{fig:2}We outline the main steps of the triad method starting with the triads constructed from the Boltzmann weights. At the end of step ($a$), we have $2d-2$ triads for a $d$-dimensional system. In the step ($b1$), we combine four copies of each triad only keeping four $x$ legs and contracting over all others. Similarly, in the step ($b2$) which is not shown, we combine them to keep four $y$ legs. The dashed color lines show the contraction along a particular direction. In step ($c1$), we fuse these four indices into two to get a matrix and perform SVD to get the projector $U$. Similarly, step ($b2$) followed by ($c2$) will give the projector $V$. Once we have $U,V$, we can take two copies of four triads and combine them with projectors as shown in step ($d$). In the next step, we can get the transformed triad after performing contractions where the order $x,y,z$ is changed to $z,x,y$. This means that we have effectively carried out the blocking in the plane orthogonal to $z$-direction. We perform the steps ($a$-$e$) two more times to do blocking along two other planes and this constitutes one full step with lattice volume increasing by eight times.}
\end{figure}
starts with the Boltzmann weight represented by a $q \times q$ real 
symmetric matrix $\mathbb{W}$ defined as: 
\begin{equation}
	\mathbb{W}_{ij}  =  \hspace{4mm} \begin{cases}
		e^{\beta} \hspace{3mm}  ; \hspace{4mm} \text{if $i = j$}   \quad \phantom{\infty}   \\
		1 \hspace{5mm}; \hspace{4mm} \text{otherwise}   \quad .\phantom{0}
	\end{cases}
\end{equation} 
To construct the four initial triads, we perform the factorization as below: 
\begin{equation}
	\mathbb{W} = SDS^{T} = \underbrace{QQ^{T}}_{\rm{Cholesky}},
\end{equation}
and the triads as:

\begin{align}
	A_{xya} &= Q_{ax}Q_{ay} \nonumber \\ 
	B_{azb} &= \mathbb{I}_{ab}Q_{az} \nonumber  \\ 
	C_{bz^{\prime}c} &= \mathbb{I}_{bc}Q_{bz^{\prime}} \nonumber  \\ 
	D_{cy^{\prime}x^{\prime}} &= Q_{cy^{\prime}}Q_{cx^{\prime}}. 
\end{align} 
The local tensor is defined in the usual way: $T_{xx^{\prime}yy^{\prime}zz^{\prime}} = A_{xya}B_{azb}C_{bz^{\prime}c}D_{cy^{\prime}x^{\prime}}$. 
In fact, for the Potts model, we never need to explicitly construct $T$ at any stage. 
We also note that while the fundamental initial tensor $T$ 
is unique for a given model, the triads can be different based on the way we have 
chosen to decompose the tensor. The optimum choice of triads for more complicated models 
remains an open problem.

\subsection{\label{subsec:1.3}Thermodynamic potentials using tensors}
The tensor network coarse-graining procedure provides an accurate evaluation of the partition function. 
Using this, we can compute the free energy, internal energy, and their derivatives. 
In order to accurately compute the critical temperature corresponding to the 
first-order phase transitions, we compute the internal energy by doing a finite-difference of the logarithm of 
partition function and then fit the data using a four-parameter sigmoidal ansatz to determine the critical temperature. The free energy density is computed as follows:
\begin{equation}
	f  =  -\frac{T}{V}\ln Z,
\end{equation}
which in tensor notation can be computed by using the tensor norm at each step of coarse-graining. 
In every coarse-graining step, we normalize the tensor by the maximum element of that tensor (which we call `norm'). We can then calculate the free energy density ($f$) 
from these normalization factors as:
\begin{equation}
	\label{eq:free1}
	f  =  -\frac{1}{\beta}\Bigg(\sum_{i=1} ^{N} 
	\frac{\log(\mathtt{norm_{i})}}{2^i} + \frac{\log(Z_{N})}{2^N}\Bigg) \
	= - \frac{1}{\beta 2^N} \Bigg(\sum_{i=1} ^{N} 
	\log(\mathtt{norm_{i}}) 2^{N-i} + \log Z_{N} \Bigg), 
\end{equation}
where $Z_{N}$ is the scalar calculated from the tensor contraction 
after the last step of coarse-graining and $2^{N}$ is the lattice volume and $N$ is the 
number of times we do coarse-graining (i.e., number of times we do step ($a$) through step ($e$) of Fig.~\ref{fig:2}).  In this paper, we have usually considered $N=24$. The internal energy density $E$ is defined as:
\begin{equation}
	E  =  \frac{T^2}{V} \frac{\partial \ln Z}{\partial T},
\end{equation}
and we compute it by taking the numerical derivative of $Z$ obtained from tensor contraction. 

\section{\label{sec:2}Results}
\subsection{\label{subsec:3.1}$q = 3 $ and $10$ - Comparison to old results} 
\begin{figure}
	\centering 
	\includegraphics[width=0.7\textwidth]{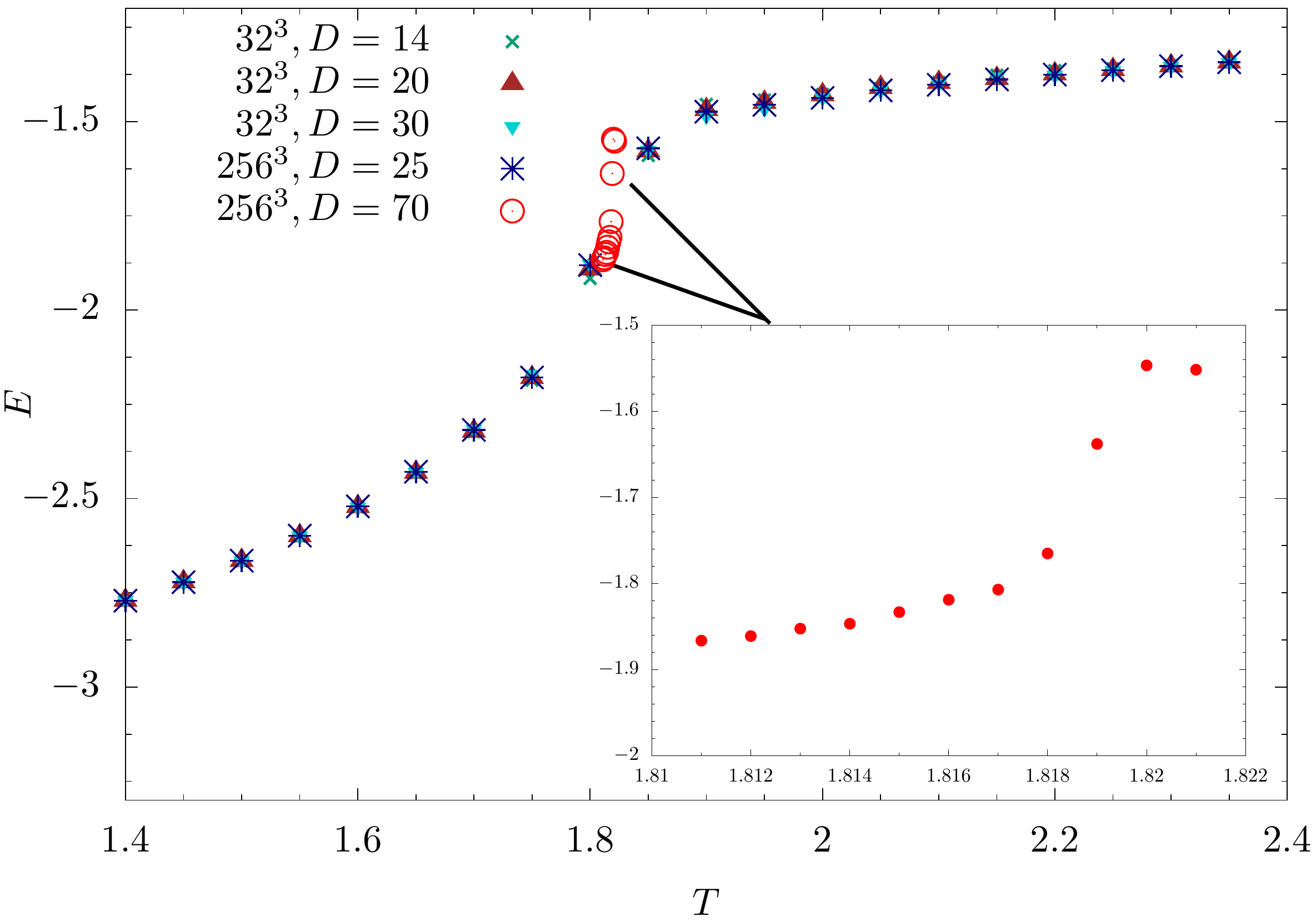}
	\caption{\label{fig:3}The dependence of $q=3$ internal energy on bond dimension $D$ and lattice volume.  The critical temperature is found to be $T_{c} = 1.8175(15)$.}
\end{figure}

The $q=3$ case has been well-studied and we use this as
a check of our computations. The extensive explorations of this model have made it clear that this case 
has a weak first-order transition and the result obtained from most recent Monte Carlo is $T_{c} = 1.81632(6)$. In the past decade, tensor network methods were only used for $q=3$ Potts model. For this case, it was found in Ref.~\cite{Wang2014} that $T_{c} = 1.8166(5)$ using the HOTRG method.  Using triads, we compute the location of phase transition to be $T_{c} = 1.8175(15)$ as seen in Fig.~\ref{fig:3}. We found that the determination of the critical temperature for $q=3$ is harder than that for large $q$. This can be seen in Fig.~\ref{fig:3} where it is not possible to fit using our ansatz. We attribute this issue to the fact that for $q=3$, the transition is only weakly first-order.  As the transition becomes strongly first-order with increasing $q$, tensor approximation method we have used works effectively to locate the critical temperature. Though the tensor network methods were not used beyond $q=3$ until this work, the model was well-studied using Monte Carlo starting already in 1970s\footnote{Ref.~\cite{Herrmann1979}, was one of the earliest papers studying $q=3$}
up to $q=6$ and we refer the reader to the old literature in Table II of Ref.~\cite{Wu1982} but the most recent extensive study of this model was carried out in Ref.~\cite{Bazavov:2008qg} for $ 3 \le q \le 10$ where critical temperatures and thermodynamic observables were computed. The most precise MC result i.e., $T_{c} = 1.816316(33)$ for $q=3$ was obtained using some new techniques in Ref.~\cite{Janke:1996qb}. 
For the $q=10$ model, the MC result locate $T_{c}$ to be 1.13446(4) \cite{Bazavov:2008qg}. 
Our results are in slight disagreement and we obtain $T_{c} = 1.1328(10)$ and 
is shown in bottom right panel of Fig.~\ref{fig:AP1}. It is not clear whether this is because of the small volume 
considered for $q=10$ in  Ref.~\cite{Bazavov:2008qg} or because of the finite-$D$ effect of our tensor computations. We have found no other work to compare for this case and hence it would be desirable to resolve this in the future. 

\subsection{\label{subsec:3.2}$q > 10$ - New results}  

\begin{figure}
	\centering 
	\includegraphics[width=0.95\textwidth]{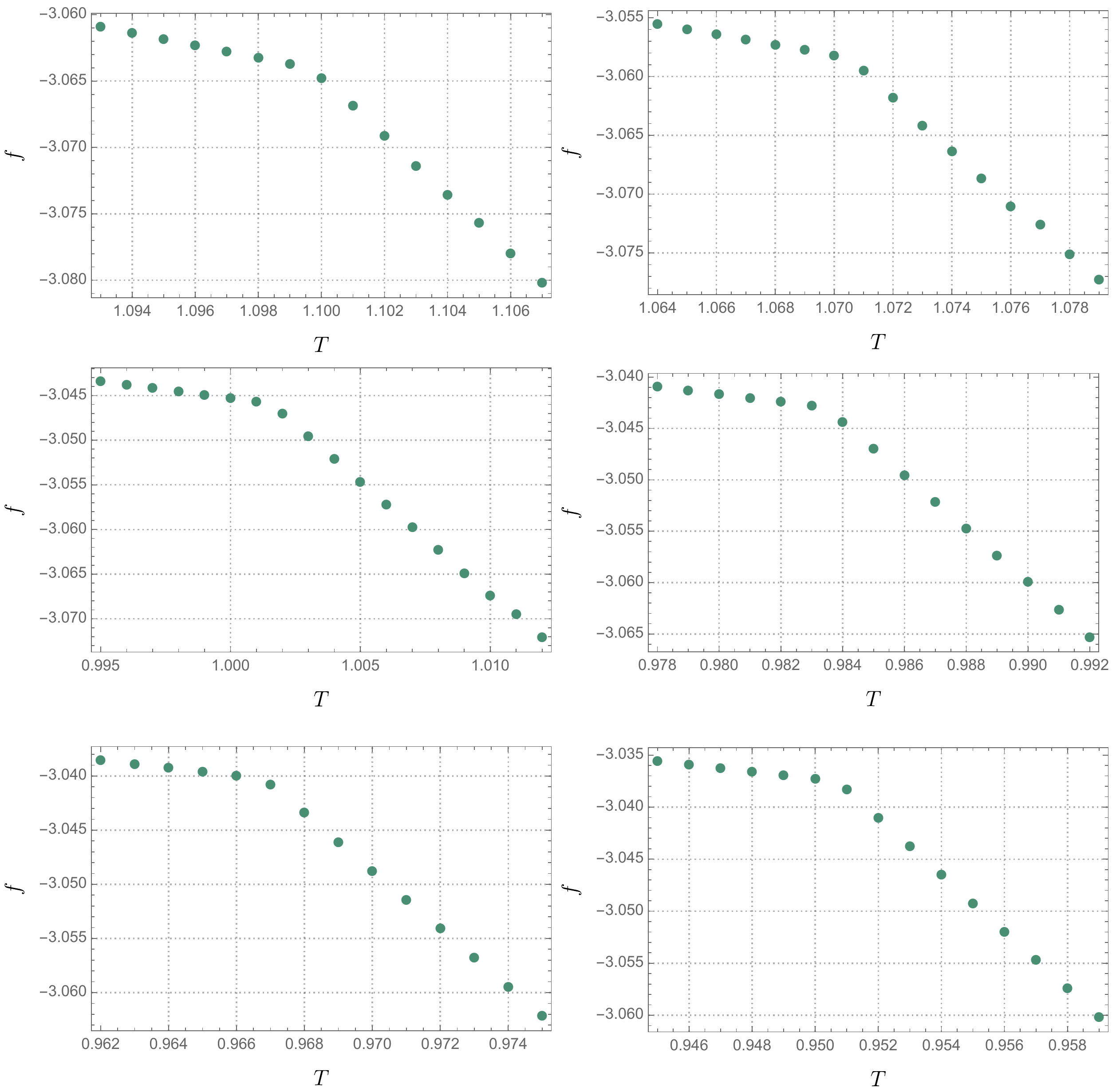}
	\caption{\label{fig:5}We show the free energy density with temperature for $q=11, 12, 15, 16, 17, 18$. The data is obtained with $D=70$ on a $256^3$ lattice.} 
\end{figure}

\begin{figure}
	\centering 
	\includegraphics[width=0.95\textwidth]{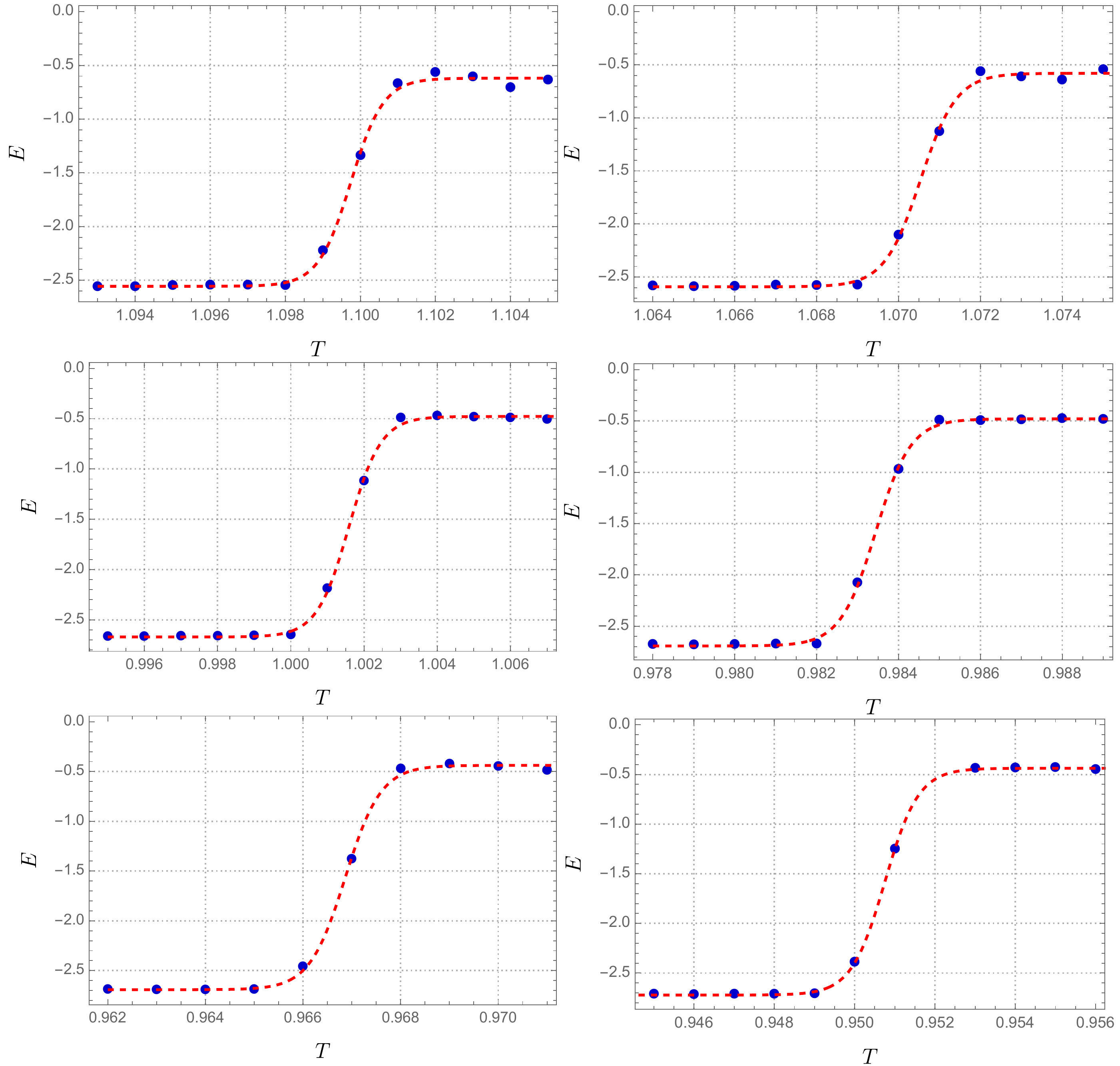}
	\caption{\label{fig:4}We show the internal energy for $q=11, 12, 15, 16, 17, 18$ with temperature. The red dashed lines represent the ansatz we used to determine the critical temperature. The data is obtained with $D=70$ on a $256^3$ lattice.}
\end{figure}

The previous subsection provides a reasonable check of the procedure we implemented using the triads.
In this subsection, we discuss new results for Potts model with $ 10 <  q < 20$. Though this method
can be effectively used to explore the large $q$ limit, we have left that for future work. For each $q$, we studied the thermodynamic potentials as shown in Fig.~\ref{fig:4} and Fig.~\ref{fig:5}  respectively. The fit to the internal energy using an ansatz gives a precise determination of the critical temperature. We collect the results for $ 10 < q \le 20$ in Table~\ref{tab:dataMC}. 
\begin{table}
\begin{center}
\begin{tabular}{ |c|c|c| } 
 \hline
 $q$ & $T_{c}$\\ 
  \hline
 11 & 1.0998(10) \\ 
 12 & 1.0705(10) \\ 
 13 & 1.0450(10) \\  
 14 & 1.0226(10) \\ 
 15 & 1.0016(10) \\ 
 16 & 0.9834(10) \\ 
 17 & 0.9669(10) \\ 
 18 & 0.9508(10) \\ 
 19 & 0.9371(10) \\ 
 20 & 0.9246(10) \\
 \hline
\end{tabular}
\caption{\label{tab:dataMC}We summarize the results for the critical temperatures obtained in this paper for $10 < q \le 20$ state Potts model.}
\end{center}
\end{table}
We used the tensor contraction package \texttt{opt\_einsum}
\cite{Smith:2018aaa} for this work. In order to obtain the results of this paper 
we make the notebook available for the reader\footnote{ \href{https://github.com/rgjha/TensorCodes/blob/master/3d/qPotts.ipynb}{\texttt{https://github.com/rgjha/TensorCodes/blob/master/3d/qPotts.ipynb}}}.
We find that the computational cost is $\mathcal{O}(D^p)$ where we obtained $p = 6.18(8)$ for the $q$-state Potts model with $q=10$ and 6.04(8) for the cubic lattice Ising model. In our computations, we used the standard SVD and not randomized SVD (RSVD) as was done in the original triad prescription. For lattice volume of $256^3$ with $D=70$, the computation for a single temperature took about 11 hours while for the same $D$ and volume of $1024^3$, it took about 14 hours on a modern laptop. This is another advantage of tensor methods: the approach to thermodynamic limit is much faster than in other numerical methods. 

\subsection{Fixed-point analysis}  
One of the main goals of the tensor renormalization group approach is to obtain the
correct fixed-point tensor. This is easier said than done and though it might be possible to achieve
this reasonably accurately in two dimensions with the developments over the past decade, 
the same cannot be said for three dimensions. 
In order to visualize this fixed-point (FP) tensor more closely in different phases and across 
a phase transition it was noticed that one of the easiest ways is to compute a specific observable 
which we have referred to as $\mathbb{X}$. This observable is 
invariant under the change of scale of the tensor i.e., $T \to \lambda T$.
This quantity is computed from a fixed-point tensor as:
\begin{equation}
	\label{eq:FPT} 
	\mathbb{X} = \frac{(\rm{Tr}~T)^2}{\rm{Tr} (T \cdot T)}, 
\end{equation}
where $T = B_{ajb} C_{bjg} D_{gfi} A_{ipa}$ is a $D \times D$ matrix which is a good approximation to the FP tensor. 
After carrying out a sufficient number of coarse-graining steps, when the initial tensor has presumably
flowed to the fixed-point tensor $T_{\star} = T$, this observable would have 
converged to an integer depending on the phase of the system. In the low-temperature symmetry breaking phase, the degeneracy is $q$ while for high-temperature ($T > T_{c}$)\footnote{In this work, we use $T$ for
both the tensor with six indices and temperature. We hope this distinction is more than obvious from the context.} it is one. There is also an additional advantage of this - it is a bonafide candidate for 
determining the critical temperature for the phase transition. 
Usually, one computes the derivatives of logarithm of partition function or magnetic susceptibility
for these determinations, which are usually prone to errors in tensor methods\footnote{Though there have been 
	some progress recently with automatic differentiation 
	\cite{Liao2019} which reduces a lot of time and makes these 
	computations faster.}, 
but computing $\mathbb{X}$ is straightforward. It was in this manner that the tensor methods were used to 
compute the critical temperature for $q=3$ in Ref.~\cite{Wang2014} 
on a cubic lattice. However, we find that the tensor approximation method we use in this work 
is not able to capture this fixed-point tensor very accurately and also the convergence 
with lattice volume and bond dimensions ($D$) is erratic. It would be interesting to explore this in more detail
and to check if $\mathbb{X}$ can be reliably computed for $q \ge 4$ Potts model in three dimensions. 
To explore this issue in more detail using the triad method, we studied the Ising model 
in the vicinity of the well-known critical temperature known through earlier numerical investigations. 
The most precise result
corresponds to a continuous transition at $ T_{c} = 4.51152469(1) $ for cubic lattice.
The RG analysis shows that in the high-temperature phase i.e., $ T\gg T_{c}$, the
observable defined in (\ref{eq:FPT}) must equal 1 and in the low-temperature phase, 
it should be $q$ i.e., 2 and it should change values at $T = T_{c}$. The triad method on a $256^3$ lattice 
shows the change of $\mathbb{X}$ happens between $ 4.5121 \le T \le 4.5122$ with $D=73$ which is different from the numerical result in the literature. This is a sign that the triad method does not correctly capture the FP behaviour with $D=73$.  The existing methods usually compute the free energy at $T=T_{c}$ and find convergence to some value but since there is no exact result for the three-dimensional Ising model yet, this determination cannot be considered rigorous. Rather, we argue that computing $\mathbb{X}$ is better 
to test their effectiveness around the critical point. In this sense, the convergence of free energy is a red herring for concluding whether correct fixed-point behaviour is obtained in a given tensor network renormalization scheme. 

\section{\label{sec:3}Discussion and Summary}

One of the motivations of this work was to understand how the critical temperature changes with $q$ for three-dimensional Potts models on a cubic lattice. In two dimensions, the critical temperature for a $q$-state model can be obtained analytically, however, no such relation is yet known for the three-dimensional model. We studied the three-dimensional model on a cubic lattice and obtained critical temperatures for different $q$. Since there are already existing results for $q \le10$, we focused on the unexplored region: $q > 10$.  
Our results fill an important gap in the literature and further clarifies that the gap i.e., $ T_{c}(q) - T_{c}(q-1)$ decreases with $q$. We find that it is comparatively difficult to resolve the transition for $q=3$ which we attribute to it being weakly first-order which is clear from the variation of internal energy with temperature close to the critical value. For larger $q$, we find it straightforward to fit a sigmoid-shape ansatz and for all $q > 3$ see a strong signal of a first-order transition. To our knowledge, this work is the first known example where tensor renormalization methods have obtained results for a spin model where no other numerical methods have been applied. We envisage that further numerical studies of the Potts model with $q \ge 10$ will be done in coming years though one needs to properly control the slowing down close to the first-order critical point which becomes stronger with increasing $q$. The second issue we addressed in the paper is the reliability of the triad approximation method to obtain the fixed-point tensor. A rigorous study of RG transformations for tensor networks in the vicinity of a critical point is still an open problem. It would be very interesting to compute critical exponents for the three-dimensional Ising model, however, we expect that the local correlations which are not very well-controlled in the triad method would first need to be addressed. There are various prescriptions for removing these but they are not yet practical in three Euclidean dimensions. A refined tensor algorithm with improved fixed-point behaviour and reasonable time complexity would be pathbreaking. It would be interesting to see if the idea of triads can be combined with additional transformations like the `disentanglers' to achieve a proper RG map. We leave these questions and many more for future explorations.

\subsection*{Acknowledgements}
We thank Bus $\chi$ for assistance during the pandemic. The author is supported by a postdoctoral fellowship at the Perimeter Institute for Theoretical Physics. Research at Perimeter Institute 
is supported in part by the Government of Canada through the Department of Innovation, 
Science and Economic Development Canada and by the Province of Ontario 
through the Ministry of Economic Development, Job Creation and Trade.
The numerical computations were done on Symmetry - Perimeter's HPC system. 

\vspace{10mm} 
\appendix
\section{Appendices}
\subsection{\label{sec:D_eff}Convergence of $Z$ and internal energy dependence on $D$ for $q=10$}
\begin{figure}
	\centering 
	\includegraphics[width=0.7\textwidth]{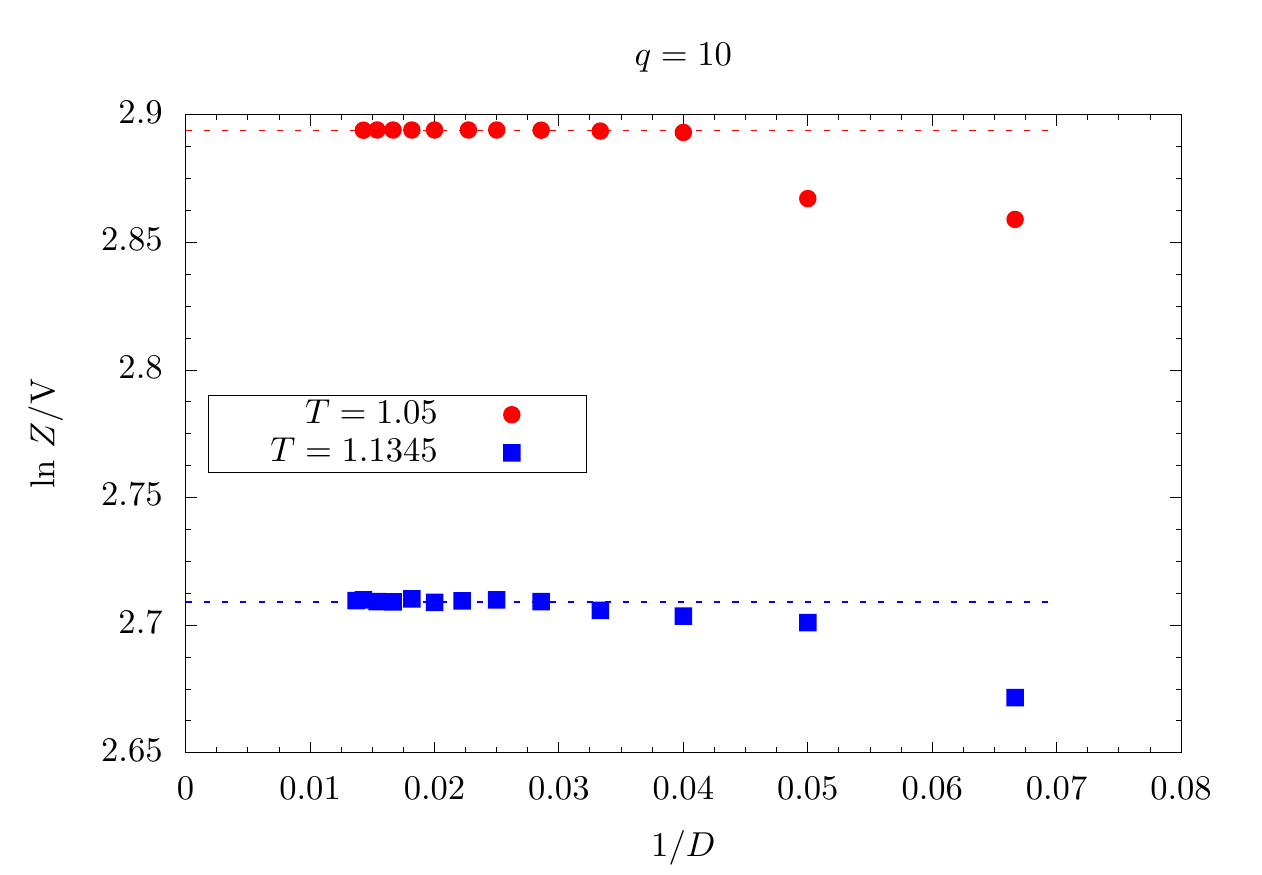}
	\caption{\label{fig:AP1} The dependence of logarithm of partition function per site for two different temperatures on cubic lattice of size $256^3$ for $q=10$. The convergence is slower around $T_{c}$. 
	}
\end{figure}

\begin{figure}
	\centering 
	\includegraphics[width=0.95\textwidth]{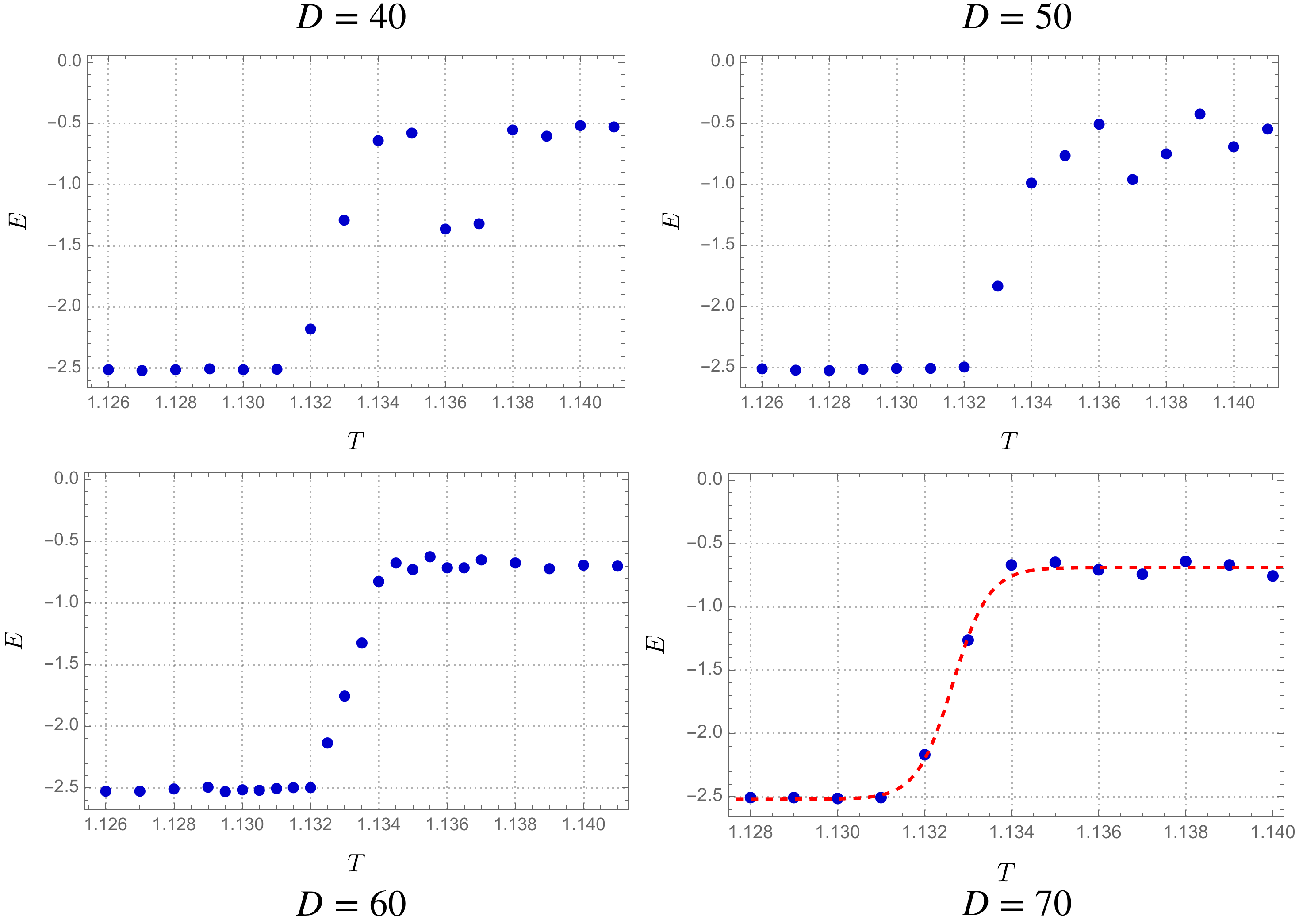}
	\caption{\label{fig:AP2}The dependence of the bond dimension $D$ on the computation of the internal energy for $q=10$. It is evident that larger $D$ improves the result. In this paper, we have 
	mostly considered $D = 70$.}
\end{figure}

In a tensor network computation, it is crucial to understand how the finite bond dimension affects the convergence of the partition function or the free energy. This convergence is a sign of how efficiently the algorithm works in practice. In cases where exact solution is available, we can compute precisely how much it deviates from the expected result. For example, it is well-known that for two-dimensional Ising model, this 
convergence is slowest around and at the critical temperature. For the $q$-state Potts model, we monitored the
convergence for two temperatures below and around $T_{c}$. Though the convergence is slow for the latter, we find that $D=70$ is a reasonable bond dimension to obtain reliable results. We show the result in 
Fig.~\ref{fig:AP1}.
We also studied the effect of the truncation of singular values at each coarse-graining step ($D$) 
on observables we compute. To understand this dependence, we studied $D = 40, 50, 60, 70$ for fixed $q=10$. 
The results are shown in Fig.~\ref{fig:AP2} where the internal energy is shown for different $D$. We see that there are some fluctuations for small $D$ which improves as we increase $D$ to 70. 

\subsection{\label{sec:V_eff}Volume dependence of internal energy for $q=15$}

\begin{figure}
	\centering 
	\includegraphics[width=0.7\textwidth]{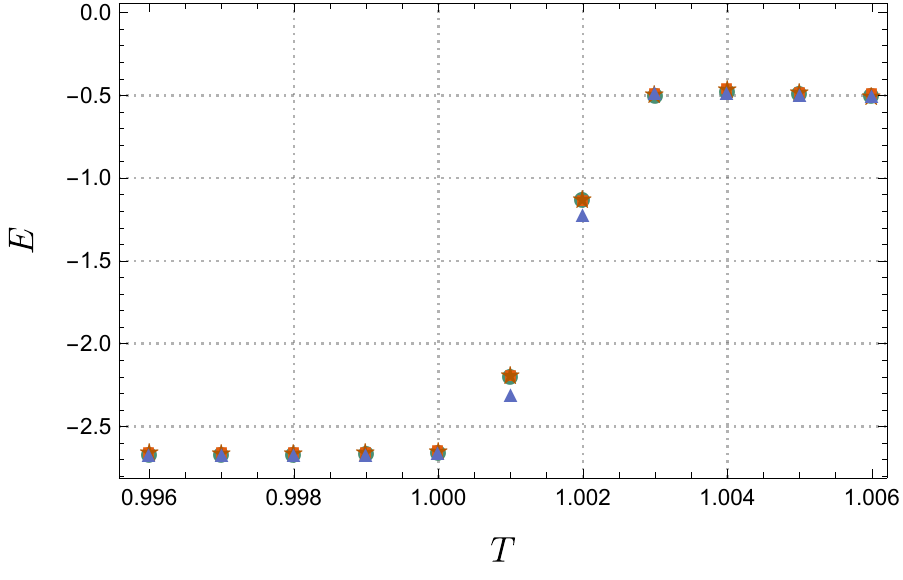}
	\caption{\label{fig:AP3} The internal energy dependence on lattice volume for $q=15$ Potts model is shown for fixed $D=70$ for $16^3$ (triangle), $32^3$ (square), $256^3$ (circle), and $1024^3$ (star) lattice volumes.} 
\end{figure}

In the previous subsection, we saw the effect of finite bond dimensions. Here, we see the finite-size effects for a
fixed $D=70$. In order to understand the finite-size effects, we compared the results obtained for different lattice volumes by fixing to $q=15$. The results obtained from $16^3, 32^3, 256^3, 1024^3$ are shown in Fig.~\ref{fig:AP3}. We find that the effects are small and ascertain that results we have obtained on a fixed lattice of $256^3$ can be correctly interpreted as thermodynamic limit. 

\bibliographystyle{scr/utphys}
\raggedright
\bibliography{bib/potts_v1.bib}

\providecommand{\href}[2]{#2}\begingroup\raggedright\begin{thebibliography}{10}

\bibitem{Gu_2009}
Z.-C. Gu and X.-G. Wen, ``Tensor-entanglement-filtering renormalization
  approach and symmetry-protected topological order,''
  \href{http://dx.doi.org/10.1103/physrevb.80.155131}{{\em Physical Review B}
  {\bfseries 80} no.~15, (Oct, 2009) }.
  \url{http://dx.doi.org/10.1103/PhysRevB.80.155131}.

\bibitem{Levin:2006jai}
M.~Levin and C.~P. Nave, ``{Tensor renormalization group approach to 2D
  classical lattice models},''
  \href{http://dx.doi.org/10.1103/PhysRevLett.99.120601}{{\em Phys. Rev. Lett.}
  {\bfseries 99} no.~12, (2007) 120601},
  \href{http://arxiv.org/abs/cond-mat/0611687}{{\ttfamily
  arXiv:cond-mat/0611687}}.

\bibitem{Xie:2009zzd}
Z.~Y. Xie, H.~C. Jiang, Q.~N. Chen, Z.~Y. Weng, and T.~Xiang, ``{Second
  Renormalization of Tensor-Network States},''
  \href{http://dx.doi.org/10.1103/PhysRevLett.103.160601}{{\em Phys. Rev.
  Lett.} {\bfseries 103} (2009) 160601},
  \href{http://arxiv.org/abs/0809.0182}{{\ttfamily arXiv:0809.0182
  [cond-mat.str-el]}}.

\bibitem{Evenbly2015}
G.~Evenbly and G.~Vidal, ``Tensor network renormalization,''
  \href{http://dx.doi.org/10.1103/physrevlett.115.180405}{{\em Physical Review
  Letters} {\bfseries 115} no.~18, (Oct., 2015) }.
  \url{https://doi.org/10.1103/physrevlett.115.180405}.

\bibitem{Xie2012}
Z.~Y. Xie, J.~Chen, M.~P. Qin, J.~W. Zhu, L.~P. Yang, and T.~Xiang,
  ``Coarse-graining renormalization by higher-order singular value
  decomposition,'' \href{http://dx.doi.org/10.1103/physrevb.86.045139}{{\em
  Physical Review B} {\bfseries 86} no.~4, (July, 2012) }.
  \url{https://doi.org/10.1103/physrevb.86.045139}.

\bibitem{Meurice:2020pxc}
Y.~Meurice, R.~Sakai, and J.~Unmuth-Yockey, ``{Tensor lattice field theory with
  applications to the renormalization group and quantum computing},''
  \href{http://arxiv.org/abs/2010.06539}{{\ttfamily arXiv:2010.06539
  [hep-lat]}}.

\bibitem{Efrati2014}
E.~Efrati, Z.~Wang, A.~Kolan, and L.~P. Kadanoff, ``Real-space renormalization
  in statistical mechanics,''
  \href{http://dx.doi.org/10.1103/revmodphys.86.647}{{\em Reviews of Modern
  Physics} {\bfseries 86} no.~2, (May, 2014) 647--667}.
  \url{https://doi.org/10.1103/revmodphys.86.647}.

\bibitem{Liu:2013nsa}
Y.~Liu, Y.~Meurice, M.~P. Qin, J.~Unmuth-Yockey, T.~Xiang, Z.~Y. Xie, J.~F. Yu,
  and H.~Zou, ``{Exact Blocking Formulas for Spin and Gauge Models},''
  \href{http://dx.doi.org/10.1103/PhysRevD.88.056005}{{\em Phys. Rev. D}
  {\bfseries 88} (2013) 056005},
  \href{http://arxiv.org/abs/1307.6543}{{\ttfamily arXiv:1307.6543 [hep-lat]}}.

\bibitem{Wang2014}
S.~Wang, Z.-Y. Xie, J.~Chen, B.~Normand, and T.~Xiang, ``Phase transitions of
  ferromagnetic potts models on the simple cubic lattice,''
  \href{http://dx.doi.org/10.1088/0256-307x/31/7/070503}{{\em Chinese Physics
  Letters} {\bfseries 31} no.~7, (July, 2014) 070503}.
  \url{https://doi.org/10.1088/0256-307x/31/7/070503}.

\bibitem{Bazavov:2019qih}
A.~Bazavov, S.~Catterall, R.~G. Jha, and J.~Unmuth-Yockey, ``{Tensor
  renormalization group study of the non-Abelian Higgs model in two
  dimensions},'' \href{http://dx.doi.org/10.1103/PhysRevD.99.114507}{{\em Phys.
  Rev. D} {\bfseries 99} no.~11, (2019) 114507},
  \href{http://arxiv.org/abs/1901.11443}{{\ttfamily arXiv:1901.11443
  [hep-lat]}}.

\bibitem{Jha:2020oik}
R.~G. Jha, ``{Critical analysis of two-dimensional classical XY model},''
  \href{http://dx.doi.org/10.1088/1742-5468/aba686}{{\em J. Stat. Mech.}
  {\bfseries 2008} (2020) 083203},
  \href{http://arxiv.org/abs/2004.06314}{{\ttfamily arXiv:2004.06314
  [hep-lat]}}.

\bibitem{Bloch:2021mjw}
J.~Bloch, R.~G. Jha, R.~Lohmayer, and M.~Meister, ``{Tensor renormalization
  group study of the three-dimensional O(2) model},''
  \href{http://dx.doi.org/10.1103/PhysRevD.104.094517}{{\em Phys. Rev. D}
  {\bfseries 104} no.~9, (2021) 094517},
  \href{http://arxiv.org/abs/2105.08066}{{\ttfamily arXiv:2105.08066
  [hep-lat]}}.

\bibitem{Hostetler:2021uml}
L.~Hostetler, J.~Zhang, R.~Sakai, J.~Unmuth-Yockey, A.~Bazavov, and Y.~Meurice,
  ``{Clock model interpolation and symmetry breaking in O(2) models},''
  \href{http://dx.doi.org/10.1103/PhysRevD.104.054505}{{\em Phys. Rev. D}
  {\bfseries 104} no.~5, (2021) 054505},
  \href{http://arxiv.org/abs/2105.10450}{{\ttfamily arXiv:2105.10450
  [hep-lat]}}.

\bibitem{Baxter1985}
R.~J. Baxter, ``Exactly solved models in statistical mechanics,''
\newblock 1982.

\bibitem{Wu1982}
F.~Y. Wu, ``The potts model,''
  \href{http://dx.doi.org/10.1103/revmodphys.54.235}{{\em Reviews of Modern
  Physics} {\bfseries 54} no.~1, (Jan., 1982) 235--268}.
  \url{https://doi.org/10.1103/revmodphys.54.235}.

\bibitem{Kadoh:2019kqk}
D.~Kadoh and K.~Nakayama, ``{Renormalization group on a triad network},''
  \href{http://arxiv.org/abs/1912.02414}{{\ttfamily arXiv:1912.02414
  [hep-lat]}}.

\bibitem{Hartmann2005}
A.~K. Hartmann, ``Calculation of partition functions by measuring component
  distributions,'' \href{http://dx.doi.org/10.1103/physrevlett.94.050601}{{\em
  Physical Review Letters} {\bfseries 94} no.~5, (Feb., 2005) }.
  \url{https://doi.org/10.1103/physrevlett.94.050601}.

\bibitem{Zhao_2010}
H.~H. Zhao, Z.~Y. Xie, Q.~N. Chen, Z.~C. Wei, J.~W. Cai, and T.~Xiang,
  ``Renormalization of tensor-network states,''
  \href{http://dx.doi.org/10.1103/physrevb.81.174411}{{\em Physical Review B}
  {\bfseries 81} no.~17, (May, 2010) }.
  \url{http://dx.doi.org/10.1103/PhysRevB.81.174411}.

\bibitem{genzor2020tensor}
J.~Genzor, T.~Nishino, and A.~Gendiar, ``Tensor networks: Phase transition
  phenomena on hyperbolic and fractal geometries,''
  \href{http://arxiv.org/abs/2003.11244}{{\ttfamily arXiv:2003.11244
  [cond-mat.stat-mech]}}.

\bibitem{Herrmann1979}
H.~J. Herrmann, ``Monte carlo simulation of the three-dimensional potts
  model,'' \href{http://dx.doi.org/10.1007/bf01321243}{{\em Zeitschrift for
  Physik B Condensed Matter and Quanta} {\bfseries 35} no.~2, (June, 1979)
  171--175}. \url{https://doi.org/10.1007/bf01321243}.

\bibitem{Bazavov:2008qg}
A.~Bazavov, B.~A. Berg, and S.~Dubey, ``{Phase Transition Properties of 3D
  Potts Models},''
  \href{http://dx.doi.org/10.1016/j.nuclphysb.2008.04.020}{{\em Nucl. Phys. B}
  {\bfseries 802} (2008) 421--434},
  \href{http://arxiv.org/abs/0804.1402}{{\ttfamily arXiv:0804.1402 [hep-lat]}}.

\bibitem{Janke:1996qb}
W.~Janke and R.~Villanova, ``{Three-dimensional three state Potts model
  revisited with new techniques},''
  \href{http://dx.doi.org/10.1016/S0550-3213(96)00710-9}{{\em Nucl. Phys. B}
  {\bfseries 489} (1997) 679--696},
  \href{http://arxiv.org/abs/hep-lat/9612008}{{\ttfamily
  arXiv:hep-lat/9612008}}.

\bibitem{Smith:2018aaa}
D.~Smith and J.~Gray, ``{opt\_einsum - A Python package for optimizing
  contraction order for einsum-like expressions},''
  \href{http://dx.doi.org/10.21105/joss.00753}{{\em Journal of Open Source
  Software} {\bfseries 3(26)} (2018) 753}.

\bibitem{Liao2019}
H.-J. Liao, J.-G. Liu, L.~Wang, and T.~Xiang, ``Differentiable programming
  tensor networks,'' \href{http://dx.doi.org/10.1103/physrevx.9.031041}{{\em
  Physical Review X} {\bfseries 9} no.~3, (Sept., 2019) }.
  \url{https://doi.org/10.1103/physrevx.9.031041}.

\end{thebibliography}\endgroup
\end{document}